\documentclass[12pt]{article}
\usepackage{amsmath}
\usepackage{amssymb}
\usepackage{fancyhdr}
\renewcommand{\maketitle}{
    \begin{center}
      \Large
        {\bf Dirac Monopole from Lorentz Symmetry in
$N$-Dimensions:~~I. The Generator Extension}
        \vskip .3 true cm
      \normalsize
        Martin Land \\
        \vskip .3 true cm
        Department of Computer Science \\
        Hadassah College \\
        P. O. Box 1114, Jerusalem 91010, Israel
      \end{center}
      \vskip .5 true cm
}

\begin{document}

\begin{flushright} HCCSP-1001-06 \end{flushright} \vskip 2 true cm

\title{}
\author{}
\maketitle

\begin{abstract}
It is by now well-known that a Lorentz force law and the homogeneous Maxwell
equations can be derived from commutation relations among Euclidean coordinates
and velocities, without explicit reference to momentum, gauge potential, action
or variational principle. More generally, it has been shown that the
specification of commutation relations in the coordinate-velocity basis
determines a unique Langrangian, from which the full canonical system follows.
This result was extended to the relativistic case and shown to correspond to a
Stueckelberg-type quantum theory, in which the underlying gauge symmetry depends
on the invariant evolution parameter, such that the associated the
five-dimensional electromagnetism becomes standard Maxwell theory in the
equilibrium limit. B\'{e}rard, Grandati, Lages and Mohrbach have studied the Lie
algebra associated with the O(3) rotational invariance of the Euclidean
coordinate-velocity system, and found an extension of the generators that
restores the commutation relations in the presence of a Maxwell field, and
renders the extended generator a constant of the classical motion.  The algebra
imposes conditions on the Maxwell field, leading to a Dirac monopole solution.
In this paper, we study the generalization of the B\'{e}rard, Grandati, Lages
and Mohrbach construction to the Lorentz generators in $N$-dimensional Minkowski
space. We find that that the construction can be maximally satisfied in a three
dimensional subspace of the full Minkowski space; this subspace can be chosen to
describe either the O(3)-invariant space sector, or an \hbox{O(2,1)-invariant}
restriction of spacetime.  The field solution reduces to the Dirac monopole
field found in the nonrelativistic case when the O(3)-invariant subspace is
selected. When an \hbox{O(2,1)-invariant} subspace is chosen, the Maxwell field can be
associated with a Coulomb-like potential of the type $A^{\mu }(x)=n^{\mu }/\rho
$, where $\rho =(x^{\mu }x_{\mu })^{1/2}$, similar to that used by Horwitz and
Arshansky to obtain a covariant generalization of the hydrogen-like bound state.
In both cases, the extended generator is conserved with respect to the invariant
parameter under classical relativistic system evolution.
\end{abstract}

\baselineskip7mm \parindent=0cm \parskip=10pt

\section{Introduction}

Since Dyson published his account \cite{dyson} of Feynman's early work on
the subject, it has become well known that posing commutation relations of
the form 
\begin{equation}
\left[ x^{i},x^{j}\right] =0\mbox{\qquad\qquad}m\,\left[ x^{i},\dot{x}^{j}%
\right] =i\hbar \,\delta ^{ij},  \label{old1}
\end{equation}%
among the quantum operators for Euclidean position and velocity, where $\dot{%
x}^{i}=dx^{i}/dt$ and $i,j=1,2,3$, restricts the admissible forces in the
classical Newton's second law 
\begin{equation}
m\ddot{x}^{i}=F^{i}(t,x,\dot{x})  \label{old2}
\end{equation}%
to the form 
\begin{equation}
m\ddot{x}^{i}=E^{i}(t,x)+\epsilon ^{ijk}\dot{x}_{j}H_{k}(t,x)  \label{old3}
\end{equation}%
with fields that must satisfy 
\begin{equation}
\nabla \cdot \mathbf{H}=0\mbox{\qquad\qquad}\nabla \times \mathbf{E}+\frac{%
\partial }{\partial t}\mathbf{H}=0.  \label{old4}
\end{equation}%
The velocity-dependent part of the interaction in (\ref{old3}) enters
through 
\begin{equation}
m^{2}\left[ \dot{x}^{i},\dot{x}^{j}\right] =-i\hbar F^{ij}(t,x)=-i\hbar
\epsilon ^{ijk}H_{k}(t,x)  \label{old5}
\end{equation}%
which is posed as a naive relaxation of assumptions about the velocity
operators, and not intended to presuppose the existence of a canonical
momentum. Although Dyson treated the \textquotedblleft
derivation\textquotedblright\ as something of a curiosity, his article led
to small flurry of new results, in particular the proof \cite{H-S} that the
assumptions (\ref{old1}) are sufficiently strong to establish the
self-adjointness of the differential equations (\ref{old2}). It follows from
self-adjointness that this system is equivalent to a unique nonrelativistic
Lagrangian mechanics \cite{Santilli} with canonical momenta whose
relationship to the velocities leads directly to (\ref{old5}). Several
authors observed \cite{conflict} that supposing Lorentz covariance in (\ref%
{old4}) conflicts with the Euclidean assumptions in (\ref{old1}), and so (%
\ref{old3}) cannot be interpreted as the Lorentz force in Maxwell theory.
These results were generalized to the relativistic case \cite{Tanimura,
beyond} in curved $N$-dimensional spacetime by taking%
\begin{equation}
\lbrack x^{\mu },x^{\nu }]=0\mbox{\qquad\qquad}m[x^{\mu },\dot{x}^{\nu
}]=-i\hbar g^{\mu \nu }(x)\mbox{\qquad\qquad}\left[ \dot{x}^{\mu },\dot{x}%
^{\nu }\right] =\left( -\dfrac{i\hbar }{m^{2}}\right) F^{\mu \nu }
\label{eqn:2.2}
\end{equation}%
and 
\begin{equation}
m\ddot{x}^{\mu }=F^{\mu }(\tau ,x,\dot{x}).  \label{eqn:2.3}
\end{equation}%
where $\mu ,\nu =0,1,\cdots ,N-1$ and $x^{\mu }(\tau )$ and its derivatives
are function of the Poincar\'{e}-invariant evolution parameter $\tau $. The
resulting system%
\begin{equation}
m[\ddot{x}^{\mu }+\Gamma ^{\mu \lambda \nu }\dot{x}_{\lambda }\dot{x}_{\nu
}]=G^{\mu }\left( \tau ,x\right) +F^{\mu \nu }\left( \tau ,x\right) \dot{x}%
_{\nu }  \label{eqn:sum1}
\end{equation}%
in which the covariant derivative contains the usual affine connection%
\begin{equation}
\Gamma _{\mu \nu \rho }=\frac{1}{2}\left( \partial _{\rho }g_{\mu \nu
}+\partial _{\nu }g_{\mu \rho }-\partial _{\mu }g_{\nu \rho }\right) 
\label{eqn:sum2}
\end{equation}%
and the fields satisfy%
\begin{equation}
\partial _{\mu }F_{\nu \rho }+\partial _{\nu }F_{\rho \mu }+\partial _{\rho
}F_{\mu \nu }=0\mbox{\qquad}\partial _{\mu }G_{\nu }-\partial _{\nu }G_{\mu
}+{\frac{\partial }{\partial \tau }F_{\mu \nu }}=0  \label{eqn:sum3}
\end{equation}%
is equivalent to the $(N+1)$-dimensional gauge theory associated with
Stueckelberg's relativistic mechanics \cite{Stueckelberg, saad}. Formally
extending the indices to $\left( N+1\right) $-dimensions%
\begin{equation}
\mu ,\nu ,\lambda =0,1,\cdots ,N-1\mbox{\qquad\qquad}\alpha ,\beta ,\gamma
=0,\cdots ,N
\end{equation}%
\begin{equation}
x^{N}=\tau \qquad \partial _{\tau }=\partial _{N}\qquad F_{\mu N}=-F_{N\mu
}=G_{\mu }  \label{eqn:2.27}
\end{equation}%
equations (\ref{eqn:sum1}) and (\ref{eqn:sum3}) become%
\begin{equation}
m[\ddot{x}^{\mu }+\Gamma ^{\mu \lambda \nu }\dot{x}_{\lambda }\dot{x}_{\nu
}]=F^{\mu \beta }(\tau ,x)\dot{x}_{\beta }  \label{eqn:2.30}
\end{equation}%
and%
\begin{equation}
\partial _{\alpha }F_{\beta \gamma }+\partial _{\beta }F_{\gamma \alpha
}+\partial _{\gamma }F_{\alpha \beta }=0.
\end{equation}%
As discussed in \cite{beyond}, the inhomogeneous source equation in the
Stueckelberg theory is%
\begin{equation}
\partial _{\beta }F^{\alpha \beta }=ej^{\alpha },
\end{equation}%
which reduces to standard Maxwell theory in an equilibrium limit (with
respect to $\tau $) of the $(4+1)$-dimensional gauge theory.

More recently \cite{BGLM}, B\'{e}rard, Grandati, Lages and Mohrbach have
studied the Lie algebra associated with the O(3) invariance of this system.
Calculating commutation relations with the angular momentum 
\begin{equation}
L_{i}=m\epsilon _{ijk}x^{i}\dot{x}^{j}  \label{BGLM1}
\end{equation}%
the noncommutivity of the velocities in (\ref{old5}) leads to
field-dependent terms,%
\begin{eqnarray}
\left[ x_{i},L_{j}\right] &=&-i\hbar \epsilon _{ijk}x_{k}  \label{BGLM2} \\
\left[ \dot{x}_{i},L_{j}\right] &=&-i\hbar \epsilon _{ijk}\dot{x}^{k}+\dfrac{%
i\hbar }{m}\delta _{ij}\left( \mathbf{x}\cdot \mathbf{B}\right) -\dfrac{%
i\hbar }{m}x_{i}B_{j}  \label{BGLM3} \\
\left[ L_{i},L_{j}\right] &=&-i\hbar \epsilon _{ijk}L^{k}-i\hbar \epsilon
_{ijk}x^{k}\left( \mathbf{x}\cdot \mathbf{B}\right) .  \label{BGLM4}
\end{eqnarray}%
The authors argue that extending the angular momentum operator $L_{i}$ to
include the O(3) invariance of the total particle-field system should
recover the closed Lie algebra. Introducing the extended angular momentum 
$\tilde{L}_{i}$ as the sum of the particle angular momentum $L_{i}$ and a
field-dependent term $Q_{i}$, 
\begin{equation}
\tilde{L}_{i}=L_{i}+Q_{i}  \label{BGLM5}
\end{equation}%
the extended commutation relations must be 
\begin{eqnarray}
\left[ x_{i},\tilde{L}_{j}\right] &=&-i\hbar \epsilon _{ijk}x_{k}
\label{BGLM6} \\
\left[ \dot{x}_{i},\tilde{L}_{j}\right] &=&-i\hbar \epsilon _{ijk}\dot{x}^{k}
\label{BGLM7} \\
\left[ \tilde{L}_{i},\tilde{L}_{j}\right] &=&-i\hbar \epsilon _{ijk}\tilde{L}%
^{k}.  \label{BGLM8}
\end{eqnarray}%
It was shown that equations (\ref{BGLM6}) to (\ref{BGLM8}) may be satisfied
with the choice%
\begin{equation}
Q_{i}=-x_{i}\left( \mathbf{x}\cdot \mathbf{B}\right) ,  \label{BGLM9}
\end{equation}%
which in turn imposes a structural condition on the field $\mathbf{B}$ given
by%
\begin{equation}
x_{j}B_{i}+x_{i}B_{j}+x_{j}x_{k}\partial _{i}B^{k}=0.  \label{BGLM10}
\end{equation}%
Since (\ref{BGLM10}) admits a solution of the form 
\begin{equation}
B_{i}=-\dfrac{x_{i}}{x^{3}}  \label{BGLM11}
\end{equation}%
the authors argue that the method has led to a magnetic monopole. Using this
solution, it is shown that the total angular momentum $\tilde{L}_{i}$ is
conserved under the classical motion.

In this paper, we generalize the B\'{e}rard, Grandati, Lages and Mohrbach
construction to the relativistic case in $N$-dimensions and study the Lie
algebra of the O($N-1$,1) generators%
\begin{equation}
M^{\mu \nu }=m\left( x^{\mu }\dot{x}^{\nu }-x^{\nu }\dot{x}^{\mu }\right) ,%
\mbox{\qquad}\mu ,\nu =0,\cdots ,N-1.  \label{gen}
\end{equation}%
On departing the realm of 3-dimensional nonrelativistic mechanics, two
immediate difficulties arise: the proliferation of terms and tensor indices,
and the conceptual difficulty of defining the magnetic monopole in $N>4$
dimensions. These difficulties are conveniently overcome in the spacetime
algebra formalism \cite{clifford, Matej}, which significantly reduces the
notational complexity, and facilitates a general discussion of the monopole
in higher dimensions \cite{duality}. The resulting relativistic construction
generalizes equations (\ref{BGLM1}) to (\ref{BGLM11}) and illuminates
important features of the symmetric structure not explicit in the
3-dimensional case. In section 2 we use the spacetime algebra formalism to
derive commutation relations for the Lorentz generators (\ref{gen}),
involving an antisymmetric tensor field $W^{\mu \nu }$ associated with the
noncommutivity of the velocities $\dot{x}^{\mu }$. In section 3, we seek
closed commutation relations for the extended generators 
\begin{equation}
\tilde{M}^{\mu \nu }=M^{\mu \nu }+Q^{\mu \nu }  \label{gen2}
\end{equation}%
and propose a choice for the field-dependent tensor operator $Q^{\mu \nu }$
that generalizes (\ref{BGLM9}). We find the structural conditions on the
field $W^{\mu \nu }$ imposed by this choice, which reduce to (\ref{BGLM10})
in the nonrelativistic limit. In section 4, it is shown that when the field $%
W^{\mu \nu }$ satisfies the structural conditions, the extended Lorentz
operator $\tilde{M}^{\mu \nu }$ is conserved under system evolution. In
section 5, solutions satisfying the structural conditions are shown to be of
the general form%
\begin{equation}
W^{\mu \nu }\left( x\right) =\frac{1}{{\left( {N-2}\right) !}}{\epsilon
^{\mu \nu \lambda _{0}\lambda _{1}\cdots \lambda _{N-3}}}F_{\lambda
_{0}\lambda _{1}\cdots \lambda _{N-3}}=\frac{1}{{\left( {N-3}\right) !}}%
\frac{{\epsilon ^{\mu \nu \lambda _{0}\lambda _{1}\cdots \lambda
_{N-3}}x_{\lambda _{0}}U_{\lambda _{1}\cdots \lambda _{N-3}}}}{{R\left(
x\right) }}  \label{gen3}
\end{equation}%
where $U^{\lambda _{1}\cdots \lambda _{N-3}}$ is a fixed antisymmetric
tensor of rank $N-3$ and $R\left( x\right) $ is a scalar radial function. In 
$N=4$, the field ${F}_{\lambda _{0}\lambda _{1}\cdots \lambda _{N-3}}$
reduces to the Li\'{e}nard-Wiechert solution for an electric charge moving
uniformly with four-velocity $U^{\mu }$, and since the Levi-Cevita dual
exchanges the electric and magnetic fields in the four-dimensional
electromagnetic field tensor, (\ref{gen3}) can be interpreted as a
generalization of the Dirac monopole found in the nonrelativistic case. This
interpretation is explored in a second paper. The structural conditions on
the fields $W^{\mu \nu }$ are shown to imply that 
\begin{equation}
x_{\lambda _{1}}U^{\lambda _{1}\cdots \lambda _{N-3}}=0,  \label{gen5}
\end{equation}%
equivalent to the requirement that $x^{\mu }$ be orthogonal to $N-3$
mutually orthogonal vectors in $N$-dimensions. Thus, the dynamical evolution 
$x^{\mu }\left( \tau \right) $ is restricted to the 3-dimensional subspace
normal to $U$, which we denote%
\begin{equation}
x^{U}=\left\{ x~|~x_{\lambda _{1}}U^{\lambda _{1}\cdots \lambda
_{N-3}}=0\right\} ,
\end{equation}%
and only the three Lorentz generators that leave $x^{U}$ invariant can be
made to satisfy closed commutation relations. Naturally, this restriction
has no consequences in the nonrelativistic case. In $N=4$, we may take the
vector $U=\hat{t}$ along the time axis, thereby recovering the
O(3)-invariant solution obtained by B\'{e}rard, Grandati, Lages and
Mohrbach, with the radial function 
\begin{equation}
R\left( x\right) =r^{3}=\left( \mathbf{x}^{2}\right) ^{3/2}
\end{equation}%
defined on the three space dimensions. On the other hand, by taking $U=\hat{n%
}$ to be spacelike, the general solution (\ref{gen3}) becomes%
\begin{equation}
W^{\mu \nu }\left( x\right) =\epsilon ^{\mu \nu \lambda \rho }\frac{\hat{n}%
_{\lambda }~x_{\rho }}{\left( x_{\perp }^{2}\right) ^{3/2}}  \label{gen6}
\end{equation}%
whose support is on the O(2,1)-invariant subspace 
\begin{equation}
x^{\hat{n}}=\left\{ x~|~x\cdot \hat{n}=0\right\} 
\end{equation}%
and the three Lorentz generators (two boosts and one rotation) that leave
this subspace invariant will satisfy the closed Lorentz algebra. The
field strength (\ref{gen3}) is associated with a potential of the type 
\begin{equation}
V\left( x\right) \sim \left( x_{\perp }^{2}\right) ^{-1/2}  \label{gen7}
\end{equation}%
which may be seen as a relativistic generalization of the nonrelativistic
Coulomb potential. A solution to the relativistic bound state problem for
the scalar hydrogen atom was found \cite{bound} in the context of the
Horwitz-Piron \cite{H-P} formalism, using a potential of the form (\ref{gen7}%
). It was shown that a discrete Schrodinger-like spectrum emerges when the
dynamics are restricted to the O(2,1)-invariant subspace%
\begin{equation}
\mathrm{RMS}\left( \hat{n}\right) =\left\{ x~|~\hat{n}^{2}>0\;,\;\left(
x\cdot \hat{n}\right) ^{2}\geq 0\right\} .
\end{equation}%
The connection between these cases is discussed in the subsequent paper.

\section{Commutation Relations}

\subsection{Spacetime algebra}

The spacetime algebra formalism \cite{clifford} achieves a high degree of
notational compactness by representing the usual tensorial objects of
physics as index-free elements in a Clifford algebra. The product of two
vectors separates naturally into a symmetric part and antisymmetric part 
\begin{equation}
ab=\tfrac{1}{2}\left( {ab+ba}\right) +\tfrac{1}{2}\left( {ab-ba}\right)
=a\cdot b+a\wedge b
\end{equation}%
where the symmetric part is identified with the scalar inner product, and
the rank 2 antisymmetric part is called a bivector. The general Clifford
number is a direct sum of multivectors of rank $0,1,\ldots ,N$ 
\begin{eqnarray}
A &=&A_{0}+A_{1}+A_{2}+A_{3}+\cdots +A_{N} \\
&=&A_{0}+A_{1}^{i}{\mathbf{e}}_{i}+\tfrac{1}{2}A_{2}^{ij}{\mathbf{e}}%
_{i}\wedge {\mathbf{e}}_{j}+\cdots +\tfrac{1}{N!}A_{N}^{i_{0}i_{2}\cdots
i_{N-1}}{\mathbf{e}}_{i_{0}}\wedge {\mathbf{e}}_{i_{1}}\wedge \cdots \wedge {%
\mathbf{e}}_{i_{N-1}}
\end{eqnarray}%
expanded on the basis 
\begin{equation}
\left\{ {1,{\mathbf{e}}_{i},{\mathbf{e}}_{i}\wedge {\mathbf{e}}_{j},{\mathbf{%
e}}_{i}\wedge {\mathbf{e}}_{j}\wedge {\mathbf{e}}_{k},\cdots ,{\mathbf{e}}%
_{0}\wedge {\mathbf{e}}_{1}\wedge \cdots \wedge {\mathbf{e}}_{N-1}}\right\}
\;\;.
\end{equation}%
The most important algebraic rules are%
\begin{eqnarray}
aA_{r} &=&a\left( {a_{1}\wedge a_{2}\wedge \cdots \wedge a_{r}}\right)
=a\cdot A_{r}+a\wedge A_{r} \\
a\cdot A_{r} &=&\sum\nolimits_{k=1}^{r}{\left( {-1}\right) ^{k+1}\left( {%
a\cdot a_{k}}\right) ~}a_{1}\wedge \cdots \wedge a_{k-1}\wedge a_{k+1}\wedge
\cdots \wedge a_{r} \\
a\wedge A_{r} &=&a\wedge a_{1}\wedge a_{2}\wedge \cdots \wedge a_{r} \\
\mathbf{i} &=&{\mathbf{e}}_{0}\wedge {\mathbf{e}}_{1}\wedge \cdots \wedge {%
\mathbf{e}}_{N-1} \\
\mathbf{i}^{2} &=&\left( {-1}\right) ^{\frac{{N\left( {N-1}\right) }}{2}%
}g_{00}\cdots g_{N-1,N-1} \\
\mathbf{i}\left[ {{\mathbf{e}}_{k_{1}}\wedge \cdots \wedge {\mathbf{e}}%
_{k_{r}}}\right]  &=&g_{k_{1}k_{1}}\cdots g_{k_{r}k_{r}}\tfrac{1}{\left(
N-r\right) {!}}{\ \epsilon ^{k_{1}\cdots k_{r}\ k_{r+1}\cdots k_{N}}\left[ {%
\ {\mathbf{e}}_{k_{r+1}}\wedge \cdots \wedge {\mathbf{e}}_{kN}}\right] } \\
a\cdot \left( \mathbf{i}{A_{r}}\right)  &=&\left( {-1}\right) ^{N-1}\mathbf{i%
}\left( {a\wedge A_{r}}\right)   \label{du-1} \\
a\wedge \left( \mathbf{i}{A_{r}}\right)  &=&\left( {-1}\right) ^{N-1}\mathbf{%
i}\left( {a\cdot A_{r}}\right)   \label{du-2}
\end{eqnarray}

\subsection{Representations and notation}

We begin with the commutation relations among position and velocity%
\begin{equation}
\left[ x^{\mu },x^{\nu }\right] =0\mbox{\qquad}\left[ x^{\mu },\dot{x}^{\nu }%
\right] =-\dfrac{i\hbar }{m}g^{\mu \nu }\mbox{\qquad}\left[ \dot{x}_{\mu
},f\left( x\right) \right] =\dfrac{i\hbar }{m}\partial _{\mu }f\left(
x\right)  \label{B1}
\end{equation}%
and the relations among velocities%
\begin{equation}
\left[ \dot{x}^{\mu },\dot{x}^{\nu }\right] =\left( -\dfrac{i\hbar }{m^{2}}%
\right) W^{\mu \nu }\left( x\right)  \label{B2}
\end{equation}%
in flat spacetime, where%
\begin{equation}
g^{\mu \nu }=\mathrm{diag}\left( -1,1,\cdots ,1\right) \mbox{\qquad}\mu ,\nu
=0,1,\cdots ,N-1.
\end{equation}%
The Lorentz generators are $M^{\mu \nu }=m\left( x^{\mu }\dot{x}^{\nu
}-x^{\nu }\dot{x}^{\mu }\right) $ as given in (\ref{gen}). We represent the
vector operators as%
\begin{equation}
x=x^{\mu }\mathbf{e}_{\mu }\mbox{\qquad}\dot{x}=\dot{x}^{\mu }\mathbf{e}%
_{\mu }
\end{equation}%
and the 2$^{nd}$rank antisymmetric tensors as bivectors%
\begin{equation}
W\left( x\right) =W^{\mu \nu }\left( x\right) \mathbf{e}_{\mu }\otimes 
\mathbf{e}_{\nu }=\frac{1}{2}W^{\mu \nu }\left( x\right) \mathbf{e}_{\mu
}\wedge \mathbf{e}_{\nu }
\end{equation}%
The entities $x$ and $\dot{x}$ are thus composed of operator-valued
components $x^{\mu }$ and $\dot{x}^{\mu }$ that are noncommuting in the
operator space but commute in the Clifford algebra, and basis vectors $%
\mathbf{e}_{\mu }$ that are noncommuting in the Clifford algebra but commute
with the operator-valued components. Exercising care with operator ordering,
the manifest antisymmetry of the Lorentz generator permits us to represent
the index-free tensor%
\begin{equation}
M=M^{\mu \nu }\mathbf{e}_{\mu }\otimes \mathbf{e}_{\nu }
\end{equation}%
as a vector product through%
\begin{equation}
M=m\left( x^{\mu }\dot{x}^{\nu }-x^{\nu }\dot{x}^{\mu }\right) \mathbf{e}%
_{\mu }\otimes \mathbf{e}_{\nu }=mx^{\mu }\dot{x}^{\nu }\left( \mathbf{e}%
_{\mu }\otimes \mathbf{e}_{\nu }-\mathbf{e}_{\nu }\otimes \mathbf{e}_{\mu
}\right) =m\left( x\wedge \dot{x}\right) .  \label{M1}
\end{equation}%
We may treat operators as Clifford scalars, by introducing auxiliary
constants 
\begin{equation}
D=D^{\mu }\mathbf{e}_{\mu }\mbox{\qquad}D^{\lambda }=g^{\lambda \mu }\mathbf{%
e}_{\mu }\mbox{\qquad}D_{2}=D^{\left( 2\right) }\wedge D^{\left( 1\right)
}=D^{\left( 2\right) \mu }D^{\left( 1\right) \nu }\mathbf{e}_{\mu }\wedge 
\mathbf{e}_{\nu }.  \label{D1}
\end{equation}%
The commutators (\ref{B1}) may be expressed as%
\begin{equation}
\left[ D\cdot x,x\right] =D_{\mu }\left[ x^{\mu },x^{\nu }\right] \mathbf{e}%
_{\nu }=0  \label{B0}
\end{equation}%
and%
\begin{equation}
\left[ D\cdot \dot{x},x\right] =D_{\mu }\left[ \dot{x}^{\mu },x^{\nu }\right]
\mathbf{e}_{\nu }=D_{\mu }\left( \dfrac{i\hbar }{m}g^{\mu \nu }\right) 
\mathbf{e}_{\nu }=\dfrac{i\hbar }{m}D.  \label{B3}
\end{equation}%
Similarly (\ref{B2}) can be written 
\begin{equation}
\left[ D\cdot \dot{x},\dot{x}\right] =D_{\mu }\left[ \dot{x}^{\mu },\dot{x}%
^{\nu }\right] \mathbf{e}_{\nu }=-\dfrac{i\hbar }{m^{2}}D_{\mu }W^{\mu \nu }%
\mathbf{e}_{\nu }=-\dfrac{i\hbar }{m^{2}}D\cdot W.  \label{B5}
\end{equation}

\subsection{Commutation relations}

Using (\ref{B0}) and (\ref{B3}) the commutation relations among generators
and position are found as 
\begin{equation}
\left[ D\cdot x,M\right] =m\left[ D\cdot x,x\wedge \dot{x}\right] =m\left[
D\cdot x,x\right] \wedge \dot{x}+mx\wedge \left[ D\cdot x,\dot{x}\right]
=-i\hbar x\wedge D.  \label{B7}
\end{equation}%
Similarly, the velocity commutators are 
\begin{eqnarray}
\left[ D\cdot \dot{x},M\right]  &=&m\left[ D\cdot \dot{x},x\wedge \dot{x}%
\right] =m\left[ D\cdot \dot{x},x\right] \wedge \dot{x}+mx\wedge \left[
D\cdot \dot{x},\dot{x}\right]   \notag \\
&=&-i\hbar \dot{x}\wedge D-\dfrac{i\hbar }{m}x\wedge \left( D\cdot W\right) .
\label{B9}
\end{eqnarray}%
The bivector equation (\ref{B9}) expresses the $\left( N-1\right) \left(
N-2\right) /2$ commutation relations between the Lorentz generators $M^{\mu
\nu }$ and the component of velocity $\dot{x}$ in the direction of the
arbitrary vector $D$. To obtain the commutators among the generators, it is
convenient to write the scalar%
\begin{equation}
D_{2}\cdot M=mD^{\left( 2\right) }\cdot \left[ D^{\left( 1\right) }\cdot
\left( x\wedge \dot{x}\right) \right] =m\left[ \left( D^{\left( 1\right)
}\cdot x\right) \left( D^{\left( 2\right) }\cdot \dot{x}\right) -\left(
D^{\left( 2\right) }\cdot x\right) \left( D^{\left( 1\right) }\cdot \dot{x}%
\right) \right] ,  \label{B11}
\end{equation}%
carefully preserving the order of $x$ and $\dot{x}$. Using (\ref{B7}) and (%
\ref{B9}) and extracting $M$, we are easily led to%
\begin{eqnarray}
\left[ D_{2}\cdot M,M\right]  &=&i\hbar \left[ D^{\left( 2\right) }\wedge
\left( D^{\left( 1\right) }\cdot M\right) -D^{\left( 1\right) }\wedge \left(
D^{\left( 2\right) }\cdot M\right) \right]   \notag \\
&&+i\hbar x\wedge \left[ \left( D^{\left( 2\right) }\cdot x\right) \left(
D^{\left( 1\right) }\cdot W\right) -\left( D^{\left( 1\right) }\cdot
x\right) \left( D^{\left( 2\right) }\cdot W\right) \right] .  \label{B10}
\end{eqnarray}%
The bivector equation (\ref{B10}) expresses the $\left( N-1\right) \left(
N-2\right) /2$ commutation relations between the Lorentz generators $M^{\mu
\nu }$ and the particular generator selected by the arbitrary vectors $%
D^{\left( 1\right) }$ and $D^{\left( 2\right) }$. This expression agrees
with the closed commutator when $W=0$.

\section{Restoring the Operator Algebra}

\subsection{Extended generators}

We seek the extended generator 
\begin{equation}
\tilde{M}=M+Q  \label{new1}
\end{equation}%
that satisfies the closed operator algebra%
\begin{equation}
\left[ D\cdot x,\tilde{M}\right] =-i\hbar x\wedge D  \label{new2}
\end{equation}%
\begin{equation}
\left[ D\cdot \dot{x},\tilde{M}\right] =-i\hbar \dot{x}\wedge D  \label{new3}
\end{equation}%
\begin{equation}
\left[ D_{2}\cdot \tilde{M},\tilde{M}\right] =i\hbar \left[ D^{\left(
2\right) }\wedge \left( D^{\left( 1\right) }\cdot \tilde{M}\right)
-D^{\left( 1\right) }\wedge \left( D^{\left( 2\right) }\cdot \tilde{M}%
\right) \right] .  \label{new4}
\end{equation}%
The actual commutation relations, equations (\ref{B7}), (\ref{B9}) and (\ref%
{B10}), then impose requirements on the form of the generator $Q$. Applying (%
\ref{B7}) to (\ref{new1}) it follows that%
\begin{equation}
\left[ D\cdot x,M+Q\right] =\left[ D\cdot x,M\right] +\left[ D\cdot x,Q%
\right] =-i\hbar x\wedge D+\left[ D\cdot x,Q\right] 
\end{equation}%
and comparison with (\ref{new2}) leads to%
\begin{equation}
\left[ D\cdot x,Q\right] =0  \label{new5}
\end{equation}%
and so that $Q$ is independent of $\dot{x}$ and its components commute among
themselves%
\begin{equation}
\frac{\partial }{\partial \dot{x}^{\mu }}Q=0\Rightarrow \left[ D_{2}\cdot Q,Q%
\right] =0.
\end{equation}%
Now from (\ref{new1}), (\ref{new3}), and (\ref{B9}) we find%
\begin{equation}
\left[ D\cdot \dot{x},M+Q\right] =-i\hbar \dot{x}\wedge D-\dfrac{i\hbar }{m}%
x\wedge \left( D\cdot W\right) +\left[ D\cdot \dot{x},Q\right] =-i\hbar \dot{%
x}\wedge D
\end{equation}%
so the new commutation relation is%
\begin{equation}
\left[ D\cdot \dot{x},\tilde{M}\right] =-i\hbar \dot{x}\wedge D-\dfrac{%
i\hbar }{m}\left[ x\wedge \left( D\cdot W\right) -\left( D\cdot \partial
\right) Q\right] ,  \label{new6}
\end{equation}%
where we have used (\ref{B1}). Comparing (\ref{new6}) with (\ref{new3})
leads to the first condition on $Q$%
\begin{equation}
\Delta _{1}=-\dfrac{i\hbar }{m}\left[ x\wedge \left( D\cdot W\right) -\left(
D\cdot \partial \right) Q\right] =0.  \label{R1}
\end{equation}%
To find the second condition on $Q$, we apply (\ref{new1}) to the LHS of (%
\ref{new4}) 
\begin{equation}
\left[ D_{2}\cdot \tilde{M},\tilde{M}\right] =\left[ D_{2}\cdot M,M\right] +%
\left[ D_{2}\cdot M,Q\right] +\left[ D_{2}\cdot Q,M\right]   \label{R3}
\end{equation}%
and to the RHS of (\ref{new4}) 
\begin{eqnarray}
\left[ D_{2}\cdot \tilde{M},\tilde{M}\right]  &=&i\hbar \left[ D^{\left(
2\right) }\wedge \left( D^{\left( 1\right) }\cdot M\right) -D^{\left(
1\right) }\wedge \left( D^{\left( 2\right) }\cdot M\right) \right]   \notag
\\
&&+i\hbar \left[ D^{\left( 2\right) }\wedge \left( D^{\left( 1\right) }\cdot
Q\right) -D^{\left( 1\right) }\wedge \left( D^{\left( 2\right) }\cdot
Q\right) \right] .
\end{eqnarray}%
Applying (\ref{B11}) to (\ref{R3}) provides 
\begin{eqnarray}
\left[ D_{2}\cdot M,Q\right] +\left[ D_{2}\cdot Q,M\right]  &=&i\hbar \left[
D^{\left( 2\right) }\wedge \left( D^{\left( 1\right) }\cdot Q\right)
-D^{\left( 1\right) }\wedge \left( D^{\left( 2\right) }\cdot Q\right) \right]
\notag \\
&&\mbox{\hspace{-.6 in}}-i\hbar x\wedge \left[ \left( D^{\left( 2\right)
}\cdot x\right) \left( D^{\left( 1\right) }\cdot W\right) -\left( D^{\left(
1\right) }\cdot x\right) \left( D^{\left( 2\right) }\cdot W\right) \right] 
\label{R2}
\end{eqnarray}%
so that combining (\ref{new4}), (\ref{B10}), and (\ref{R2}) leads to%
\begin{eqnarray}
\left[ D_{2}\cdot \tilde{M},\tilde{M}\right]  &=&i\hbar \left[ D^{\left(
2\right) }\wedge \left( D^{\left( 1\right) }\cdot \tilde{M}\right)
-D^{\left( 1\right) }\wedge \left( D^{\left( 2\right) }\cdot \tilde{M}%
\right) \right]   \notag \\
&&-i\hbar x\wedge \left[ \left( D^{\left( 2\right) }\cdot x\right) \left(
D^{\left( 1\right) }\cdot W\right) -\left( D^{\left( 1\right) }\cdot
x\right) \left( D^{\left( 2\right) }\cdot W\right) \right]   \notag \\
&&-i\hbar \left[ D^{\left( 2\right) }\wedge \left( D^{\left( 1\right) }\cdot
Q\right) -D^{\left( 1\right) }\wedge \left( D^{\left( 2\right) }\cdot
Q\right) \right] ,  \label{new7}
\end{eqnarray}%
and we arrive at the second condition on the generator $Q$,%
\begin{eqnarray}
\Delta _{2} &=&-i\hbar 
\Bigl%
\{x\wedge \left[ \left( D^{\left( 2\right) }\cdot x\right) \left( D^{\left(
1\right) }\cdot W\right) -\left( D^{\left( 1\right) }\cdot x\right) \left(
D^{\left( 2\right) }\cdot W\right) \right]   \notag \\
&&\mbox{\qquad\qquad\qquad}+\left[ D^{\left( 2\right) }\wedge \left(
D^{\left( 1\right) }\cdot Q\right) -D^{\left( 1\right) }\wedge \left(
D^{\left( 2\right) }\cdot Q\right) \right] 
\Bigr%
\}=0.  \label{R6}
\end{eqnarray}

\subsection{Choice of generator}

Since $Q=Q\left( x\right) $ cannot depend on $\dot{x}$ and since $D=\left(
D\cdot \partial \right) x$ allows us to write (\ref{R1}) as%
\begin{equation}
\left( D\cdot \partial \right) Q=x\wedge \left[ \left( D\cdot \partial
\right) x\right] \cdot W,
\end{equation}%
we are led to consider the form%
\begin{equation}
Q=x\wedge \left( x\cdot W\right) -x^{2}W.  \label{Q1}
\end{equation}%
Using%
\begin{equation}
x^{2}W=x\cdot \left( x\wedge W\right) +x\wedge \left( x\cdot W\right)
\label{AA0}
\end{equation}%
equation (\ref{Q1}) can be rewritten as%
\begin{equation}
Q=x\wedge \left( x\cdot W\right) -\left[ x\cdot \left( x\wedge W\right)
+x\wedge \left( x\cdot W\right) \right] =-x\cdot \left( x\wedge W\right) .
\label{Q3}
\end{equation}%
The geometrical meaning of (\ref{Q3}) can be seen by re-writing (\ref{AA0})
as%
\begin{equation}
W=\frac{1}{x^{2}}\left[ x\cdot \left( x\wedge W\right) +x\wedge \left(
x\cdot W\right) \right] =\hat{x}\cdot \left( \hat{x}\wedge W\right) +\hat{x}%
\wedge \left( \hat{x}\cdot W\right) .  \label{AA}
\end{equation}%
Since $\hat{x}\cdot W=0$ if the unit vector $\hat{x}$ is orthogonal to the
plane spanned by $W$, we find that (\ref{AA}) represents the decomposition
of $W$ into%
\begin{equation}
W_{\shortparallel }=\hat{x}\wedge \left( \hat{x}\cdot W\right) \mbox{\qquad}%
W_{\perp }=\hat{x}\cdot \left( \hat{x}\wedge W\right)  \label{AA2}
\end{equation}%
and so comparing with (\ref{Q3}), $Q$ can be described as the component of $%
W $ orthogonal to the observation point $x$.

\subsection{Conditions on the field $W$}

Applying the first condition (\ref{R1}) on the generator $Q$ to the form (%
\ref{Q3}), we find a corresponding condition on the field $W$%
\begin{equation}
\Delta _{1}=-\dfrac{i\hbar }{m}\left\{ x\wedge \left( D\cdot W\right)
+D\cdot \left( x\wedge W\right) +x\cdot \left( D\wedge W\right) +x\cdot %
\left[ x\wedge \left( D\cdot \partial \right) W\right] \right\} =0.
\label{F1}
\end{equation}%
The second condition (\ref{R6}) on the generator $Q$ is purely algebraic%
\begin{eqnarray}
\Delta _{2} &=&i\hbar \left( D^{\left( 1\right) }\cdot x\right) \left[
x\wedge \left( D^{\left( 2\right) }\cdot W\right) -D^{\left( 2\right)
}\wedge \left( x\cdot W\right) \right]  \notag \\
&&+i\hbar \left( D^{\left( 2\right) }\cdot x\right) \left[ D^{\left(
1\right) }\wedge \left( x\cdot W\right) -x\wedge \left( D^{\left( 1\right)
}\cdot W\right) \right]  \notag \\
&&+i\hbar x^{2}\left[ D^{\left( 2\right) }\wedge \left( D^{\left( 1\right)
}\cdot W\right) -D^{\left( 1\right) }\wedge \left( D^{\left( 2\right) }\cdot
W\right) \right]  \notag \\
&&+i\hbar \left( D^{\left( 2\right) }\wedge x\right) \left( \left( D^{\left(
1\right) }\wedge x\right) \cdot W\right) -i\hbar \left( D^{\left( 1\right)
}\wedge x\right) \left( \left( D^{\left( 2\right) }\wedge x\right) \cdot
W\right) =0.  \label{F2}
\end{eqnarray}

\section{Conserved Evolution}

The derivative of the classical extended Lorentz generator with respect to
the invariant time is%
\begin{eqnarray}
\dfrac{d}{d\tau }\tilde{M} &=&\dot{M}+\dot{Q}  \notag \\
&=&m\dfrac{d}{d\tau }\left( x\wedge \dot{x}\right) +\dfrac{d}{d\tau }\left[
-x\cdot \left( x\wedge W\right) \right]  \notag \\
&=&m\left( x\wedge \ddot{x}+\dot{x}\wedge \dot{x}\right) -\dot{x}\cdot
\left( x\wedge W\right) -x\cdot \left( \dot{x}\wedge W\right) -x\cdot \left(
x\wedge \dot{W}\right)  \notag \\
&=&m\left( x\wedge \ddot{x}\right) -\dot{x}\cdot \left( x\wedge W\right)
-x\cdot \left( \dot{x}\wedge W\right) -x\cdot \left( x\wedge \dot{W}\right) .
\end{eqnarray}%
Using the equations of motion for $\tau $-independent fields%
\begin{equation}
W^{\mu \nu }\left( x,\tau \right) =W^{\mu \nu }\left( x\right) \mbox{\qquad}%
W^{\mu N}\left( x,\tau \right) =0
\end{equation}%
leads to%
\begin{equation}
m\ddot{x}^{\mu }=W^{\mu \nu }\dot{x}_{\nu }=-\dot{x}_{\nu }W^{\nu \mu }%
\mbox{\qquad}\longrightarrow \mbox{\qquad}\ddot{x}=-\frac{1}{m}\dot{x}\cdot W
\end{equation}%
and since $W$ only depends on $\tau $ through $x\left( \tau \right) $, 
\begin{equation}
\dot{W}^{\mu \nu }=\partial _{\lambda }W^{\mu \nu }\dot{x}^{\lambda }%
\mbox{\qquad}\longrightarrow \mbox{\qquad}\dot{W}=\left( \dot{x}\cdot
\partial \right) W  \label{div}
\end{equation}%
the equations of motion become%
\begin{equation*}
\dfrac{d}{d\tau }\tilde{M}=-x\wedge \left( \dot{x}\cdot W\right) -\dot{x}%
\cdot \left( x\wedge W\right) -x\cdot \left( \dot{x}\wedge W\right) -x\cdot
\left( x\wedge \left( \dot{x}\cdot \partial \right) W\right) .
\end{equation*}%
Applying (\ref{F1}) with $D=\dot{x}$,%
\begin{equation}
-x\cdot \left[ x\wedge \left( \dot{x}\cdot \partial \right) W\right]
=x\wedge \left( \dot{x}\cdot W\right) +\dot{x}\cdot \left( x\wedge W\right)
+x\cdot \left( \dot{x}\wedge W\right)
\end{equation}%
we find%
\begin{eqnarray}
\dfrac{d}{d\tau }\tilde{M} &=&-x\wedge \left( \dot{x}\cdot W\right) -\dot{x}%
\cdot \left( x\wedge W\right) -x\cdot \left( \dot{x}\wedge W\right)  \notag
\\
&&+x\wedge \left( \dot{x}\cdot W\right) +\dot{x}\cdot \left( x\wedge
W\right) +x\cdot \left( \dot{x}\wedge W\right)  \notag \\
&=&0
\end{eqnarray}%
so that the classical generator is conserved.

\section{Solutions}

In the previous sections we have shown that in principle we may define an
extended Lorentz generator that is conserved under evolution of the
particle-field system and satisfies the closed O($N-1,1$) commutation
relations. The resulting system depends, of course, on the existence of a
field $W$ satisfying conditions (\ref{F1}) and (\ref{F2}). It turns out such
solutions exist in a restricted regime. We first obtain the Li\'{e}%
nard-Wiechert field for a uniformly moving charge, and demonstrate that no
generalization of this form can satisfy the condition (\ref{F1}). We then
show that under limited circumstances, the dual of this solution can satisfy
the conditions.

\subsection{Li\'{e}nard-Wiechert field in 4-dimensions}

A charge moving uniformly with four-velocity $U^{\mu }$ induces a potential $%
A^{\mu }\left( x\right) $ through the Maxwell Green's function%
\begin{equation}
A^{\mu }(x)=\frac{1}{2\pi }\int d^{4}x^{\prime }J^{\mu }\left( x^{\prime
}\right) \delta \left[ \left( x-x^{\prime }\right) ^{2}\right]  \label{LW1}
\end{equation}%
where%
\begin{equation}
J^{\mu }\left( x^{\prime }\right) =\int d\tau \;U^{\mu }\delta ^{4}\left(
x^{\prime }-U\tau \right) .  \label{curr}
\end{equation}%
Expanding the Green's function using%
\begin{equation}
\delta \left( f\left( x\right) \right) =\sum_{i}\frac{\delta \left(
x-x_{i}\right) }{\left\vert \frac{df}{dx}\right\vert _{x=x_{i}}}
\end{equation}%
we arrive at the Li\'{e}nard-Wiechert potential for uniform motion%
\begin{equation}
A^{\mu }(x)=\frac{U^{\mu }}{\left[ x^{2}+\left( x\cdot U\right) ^{2}\right]
^{1/2}}  \label{LW2}
\end{equation}%
from which we derive the field strength tensor as%
\begin{equation}
F^{\mu \nu }\left( x\right) =\frac{U^{\mu }x^{\nu }-U^{\nu }x^{\mu }}{\left[
x^{2}+\left( x\cdot U\right) ^{2}\right] ^{3/2}}.  \label{LW3}
\end{equation}%
Since the four-velocity is timelike $U^{2}=-1$, and the observation point $x$
can be resolved as%
\begin{equation}
x=-U^{2}x=-U\left( U\cdot x+U\wedge x\right)
\end{equation}%
with%
\begin{eqnarray}
x_{\shortparallel } &=&-U\left( U\cdot x\right) \\
x_{\perp } &=&-U\left( U\wedge x\right) =-U^{2}x+U\left( U\cdot x\right)
=x+U\left( U\cdot x\right) ,
\end{eqnarray}%
we recognize%
\begin{equation}
\left( x_{\perp }\right) ^{2}=x^{2}+U^{2}\left( U\cdot x\right) ^{2}+2\left(
U\cdot x\right) ^{2}=x^{2}+\left( U\cdot x\right) ^{2}.
\end{equation}%
In the index-free notation of spacetime algebra, we may rewrite (\ref{LW2})\
and (\ref{LW3}) as%
\begin{equation}
A\left( x\right) =\frac{U}{\left( x_{\perp }^{2}\right) ^{1/2}}  \label{LW4}
\end{equation}%
\begin{equation}
F\left( x\right) =\partial \wedge A(x)=\frac{U\wedge x_{\perp }}{\left(
x_{\perp }^{2}\right) ^{3/2}}=\frac{U\wedge x}{\left( x_{\perp }^{2}\right)
^{3/2}}  \label{LW5}
\end{equation}%
where we have taken advantage of the identity $U\wedge x_{\shortparallel }=0$%
.

\subsection{Electric field in $N$-dimensions}

We attempt a general solution of the form%
\begin{equation}
W\left( x\right) =\frac{U\wedge x}{R\left( x\right) }  \label{S2}
\end{equation}%
with an arbitrary fixed vector $U$. The first condition (\ref{F1}) on the
field is 
\begin{equation}
x\wedge \left( D\cdot W\right) +D\cdot \left( x\wedge W\right) +x\cdot
\left( D\wedge W\right) +x\cdot \left[ x\wedge \left( D\cdot \partial
\right) W\right] =0,  \label{S1}
\end{equation}%
in which the derivative term requires%
\begin{equation}
\left( D\cdot \partial \right) W=\left( D\cdot \partial \right) \frac{%
U\wedge x}{R\left( x\right) }=\frac{U\wedge D}{R\left( x\right) }-\frac{%
\left( U\wedge x\right) \left( D\cdot \partial \right) R\left( x\right) }{%
R^{2}\left( x\right) }
\end{equation}%
so that%
\begin{eqnarray}
x\cdot \left[ x\wedge \left( D\cdot \partial \right) W\right] &=&x\cdot %
\left[ \left( \frac{x\wedge U\wedge D}{R\left( x\right) }-\frac{\left(
x\wedge U\wedge x\right) \left( D\cdot \partial \right) R\left( x\right) }{%
R^{2}\left( x\right) }\right) \right]  \notag \\
&=&-x\cdot \left( \frac{D\wedge U\wedge x}{R\left( x\right) }\right) ,
\label{S3}
\end{eqnarray}%
where we used $x\wedge x=0$. Similarly, $D\cdot \left( x\wedge W\right) $
vanishes identically for this solution, so (\ref{S1}) becomes%
\begin{eqnarray}
0 &=&x\wedge \left( D\cdot \frac{U\wedge x}{R\left( x\right) }\right)
+x\cdot \left( \frac{D\wedge U\wedge x}{R\left( x\right) }\right) -x\cdot
\left( \frac{D\wedge U\wedge x}{R\left( x\right) }\right)  \notag \\
&=&\frac{x\wedge \left( D\cdot \left( U\wedge x\right) \right) }{R\left(
x\right) }  \notag \\
&=&\frac{x\wedge \left( \left( D\cdot U\right) x-U\left( D\cdot x\right)
\right) }{R\left( x\right) }  \notag \\
&=&\frac{\left( U\wedge x\right) \left( D\cdot x\right) }{R\left( x\right) }
\notag \\
&=&\left( D\cdot x\right) W\left( x\right) .  \label{S4}
\end{eqnarray}%
Notice that $D$ is arbitrary, specifying the component of velocity being
commuted with the generator. Since $x\wedge x\equiv 0$ causes the second
term of (\ref{S3}) to vanish for algebraic reasons, the trial solution (\ref%
{S2}) fails for any form of the radial function $R\left( x\right) $.
Moreover, expression (\ref{S2}) is seen to be an unlikely candidate for the
field solution, because it leads to%
\begin{equation}
Q=-x\cdot \left( x\wedge W\right) =-x\cdot \left( x\wedge \frac{U\wedge x}{%
R\left( x\right) }\right) \equiv 0.
\end{equation}

\subsection{Dual field in N-dimensions}

Generalizing the 3-dimensional result, we assume that the field $W\left(
x\right) $ is given as the dual of some field $F\left( x\right) $, so that
we replace 
\begin{equation}
W\left( x\right) =\mathbf{i}F\left( x\right)   \label{replacement}
\end{equation}%
in the requirement (\ref{F1}), and adapt it in $N$-dimensions by using (\ref%
{du-1}) and (\ref{du-2}). The first condition on the field $F$ then becomes 
\begin{eqnarray}
0 &=&\Delta _{1}=-\dfrac{i\hbar }{m}%
\Bigl%
\{x\wedge \left( D\cdot \mathbf{i}F\right) +D\cdot \left( x\wedge \mathbf{i}%
F\right) +x\cdot \left( D\wedge \mathbf{i}F\right) +x\cdot \left[ x\wedge
\left( D\cdot \partial \right) \mathbf{i}F\right] 
\Bigr%
\}  \notag \\
&=&-\dfrac{i\hbar }{m}\mathbf{i}%
\Bigl%
\{x\cdot \left( D\wedge F\right) +D\wedge \left( x\cdot F\right) +x\wedge
\left( D\cdot F\right) +x\wedge \left[ x\cdot \left( D\cdot \partial \right)
F\right] 
\Bigr%
\}.  \label{G1}
\end{eqnarray}%
Since the field $W$ was introduced in (\ref{B2}) as a bivector, the
Levi-Cevita dual field $F$ must be an $\left( N-2\right) $-vector in $N$%
-dimensions. Generalizing expression (\ref{S2}), we may take the general
solution for $F\left( x\right) $ to be%
\begin{equation}
F\left( x\right) =\frac{x\wedge U}{R\left( x\right) }  \label{G3}
\end{equation}%
where $U$ is a fixed multivector of rank $N-3$. Now%
\begin{equation}
\left( D\cdot \partial \right) F=\left( D\cdot \partial \right) \frac{%
x\wedge U}{R\left( x\right) }=\frac{D\wedge U}{R\left( x\right) }-\frac{%
\left( x\wedge U\right) \left( D\cdot \partial \right) R\left( x\right) }{%
R^{2}\left( x\right) }
\end{equation}%
and we have%
\begin{equation}
x\wedge \left[ x\cdot \left( D\cdot \partial \right) F\right] =\frac{\left(
x\cdot D\right) \left( x\wedge U\right) -x\wedge D\wedge \left( x\cdot
U\right) }{R\left( x\right) }-\frac{x^{2}\left( x\wedge U\right) \left(
D\cdot \partial \right) R\left( x\right) }{R^{2}\left( x\right) }.
\end{equation}%
The other terms in (\ref{G1}) are%
\begin{eqnarray}
x\wedge \left( D\cdot F\right)  &=&\frac{\left( D\cdot x\right) \left(
x\wedge U\right) }{R\left( x\right) } \\
D\wedge \left( x\cdot F\right)  &=&\frac{x^{2}D\wedge U-D\wedge x\wedge
\left( x\cdot U\right) }{R\left( x\right) } \\
x\cdot \left( D\wedge F\right)  &=&\frac{\left( x\cdot D\right) \left(
x\wedge U\right) -x^{2}D\wedge U+D\wedge x\wedge \left( x\cdot U\right) }{%
R\left( x\right) }.
\end{eqnarray}%
Assembling the terms we find%
\begin{equation}
0=\Delta _{1}=-\dfrac{i\hbar }{m}\mathbf{i}\left\{ \frac{\left( x\cdot
D\right) \left( x\wedge U\right) }{R\left( x\right) }-\frac{x^{2}\left(
x\wedge U\right) \left( D\cdot \partial \right) R\left( x\right) }{%
R^{2}\left( x\right) }-\frac{x\wedge D\wedge \left( x\cdot U\right) }{%
R\left( x\right) }\right\} .
\end{equation}%
Taking the radial function $R\left( x\right) $ to be%
\begin{equation}
R\left( x\right) =\left( x^{2}\right) ^{3/2}  \label{rad}
\end{equation}%
so that%
\begin{equation}
\frac{\left( D\cdot \partial \right) R\left( x\right) }{R^{2}\left( x\right) 
}=3\frac{D\cdot x}{x^{2}R\left( x\right) }
\end{equation}%
leads to%
\begin{equation}
0=\Delta _{1}=-\dfrac{i\hbar }{m}\ \mathbf{i}\left[ \frac{D\wedge x\wedge
\left( x\cdot U\right) }{R\left( x\right) }\right] =-\dfrac{i\hbar }{m}\ 
\mathbf{i\ }\frac{D\wedge U_{\shortparallel }}{\left( x^{2}\right) ^{1/2}}
\label{G2}
\end{equation}%
where we used (\ref{AA2}) to write 
\begin{equation}
x\wedge \left( x\cdot U\right) =x^{2}U_{\shortparallel }.
\end{equation}%
Again, since the vector $D$ is arbitrary, the condition imposed by (\ref{G2}%
) is satisfied when the dynamical evolution $x^{\mu }\left( \tau \right) $
is restricted to the subspace defined by%
\begin{equation}
x\left( \tau \right) \in x^{U}=\left\{ x~|~x\cdot U=0\right\} ,
\end{equation}%
equivalent to the requirement that $x^{\mu }$ be orthogonal to $N-3$
mutually orthogonal vectors in $N$-dimensions. We conclude that the B\'{e}%
rard, Grandati, Lages and Mohrbach construction may be generalized to any
3-dimensional subspace of $N$-dimensional Minkowski space. The generators of
the O(2,1) or O(3) subgroup of O($N-1$,1) that leave the subspace $x^{U}$
invariant have an extension that is dynamically conserved and satisfies
closed commutation relations.

The second condition on the field $W\left( x\right) $ was given in (\ref{F2}%
). With the replacement (\ref{replacement}) this condition becomes%
\begin{eqnarray}
\Delta _{2} &=&i\hbar \left( D^{\left( 1\right) }\cdot x\right) \left[ 
\mathbf{i}x\cdot \left( D^{\left( 2\right) }\wedge F\right) -\mathbf{i}%
D^{\left( 2\right) }\cdot \left( x\wedge F\right) \right]  \notag \\
&&+i\hbar \left( D^{\left( 2\right) }\cdot x\right) \left[ \mathbf{i}%
D^{\left( 1\right) }\cdot \left( x\wedge F\right) -\mathbf{i}x\cdot \left(
D^{\left( 1\right) }\wedge F\right) \right]  \notag \\
&&+i\hbar x^{2}\left[ \mathbf{i}D^{\left( 2\right) }\cdot \left( D^{\left(
1\right) }\wedge F\right) -\mathbf{i}D^{\left( 1\right) }\cdot \left(
D^{\left( 2\right) }\wedge F\right) \right]  \notag \\
&&+i\hbar \left( D^{\left( 2\right) }\wedge x\right) \left( \mathbf{i}%
D^{\left( 1\right) }\wedge \left( x\wedge F\right) \right) -i\hbar \left(
D^{\left( 1\right) }\wedge x\right) \left( \mathbf{i}D^{\left( 2\right)
}\wedge \left( x\wedge F\right) \right)
\end{eqnarray}%
Applying (\ref{G3}) and using $x\wedge x\equiv 0$, this reduces to%
\begin{eqnarray}
\Delta _{2} &=&i\hbar \frac{\mathbf{i}}{R\left( x\right) }\left( D^{\left(
1\right) }\cdot x\right) \left[ x\cdot \left( D^{\left( 2\right) }\wedge
x\wedge U\right) \right] -i\hbar \left( D^{\left( 2\right) }\cdot x\right) %
\left[ x\cdot \left( D^{\left( 1\right) }\wedge x\wedge U\right) \right] 
\notag \\
&&+i\hbar \frac{\mathbf{i}x^{2}}{R\left( x\right) }\left[ D^{\left( 2\right)
}\cdot \left( D^{\left( 1\right) }\wedge x\wedge U\right) -D^{\left(
1\right) }\cdot \left( D^{\left( 2\right) }\wedge x\wedge U\right) \right] ,
\end{eqnarray}%
so that expanding the inner products and restricting the dynamics to the
subspace \linebreak \mbox{$x^{U}=\left\{ x~|~x\cdot U=0\right\} $} leads to 
\begin{equation}
\Delta _{2}=i\hbar \frac{\mathbf{i}x^{2}}{R\left( x\right) }\left[ \left(
D^{\left( 1\right) }\wedge x\right) \wedge \left( D^{\left( 2\right) }\cdot
U\right) -\left( D^{\left( 2\right) }\wedge x\right) \wedge \left( D^{\left(
1\right) }\cdot U\right) \right] .  \label{finally2}
\end{equation}%
The arbitrary vectors $D^{\left( 1\right) }$ and $D^{\left( 2\right) }$
specify components of the Lorentz generator that undergo commutation with
the index-free Lorentz generator tensor. Since (\ref{G2}) restricts the
dynamical components to the subspace $x^{U}=\left\{ x~|~x\cdot U=0\right\} $
we may limit attention to the components for which 
\mbox{$D^{\left( 1\right)
}\cdot U=D^{\left( 2\right) }\cdot U=0$}, and so (\ref{finally2}) is
automatically satisfied.

\subsubsection{Dual field in 4-dimensions}

In $N=4$, the multivector $U$ is just a four-vector and we recognize (\ref%
{G3}) as a field of the Li\'{e}nard-Wiechert type 
\begin{equation}
F\left( x\right) =\frac{x\wedge U}{\left( x^{2}\right) ^{3/2}}\mbox{\qquad}%
\rightarrow \mbox{\qquad}W\left( x\right) =\mathbf{i}\frac{x\wedge U}{\left(
x^{2}\right) ^{3/2}}  \label{4-3}
\end{equation}%
valid in the subspace for which 
\begin{equation}
x^{U}=\left\{ x~|~x\cdot U=0\right\} .  \label{4-2}
\end{equation}%
We may recover the nonrelativistic case by choosing the unit vector along
the time axis $U=\mathbf{e}_{0}$, imposing the restriction of the dynamics to%
\begin{equation}
x\left( \tau \right) =\left( 0,\mathbf{x}\right)   \label{4-1}
\end{equation}%
so that 
\begin{equation}
F\left( x\right) =\frac{\mathbf{e}_{0}\wedge \mathbf{x}}{r^{3}}=\frac{x^{1}}{%
r^{3}}\mathbf{e}_{0}\wedge \mathbf{e}_{1}+\frac{x^{2}}{r^{3}}\mathbf{e}%
_{0}\wedge \mathbf{e}_{2}+\frac{x^{3}}{r^{3}}\mathbf{e}_{0}\wedge \mathbf{e}%
_{3}
\end{equation}%
and 
\begin{equation}
W\left( x\right) =E^{i}\mathbf{e}_{0}\wedge \mathbf{e}_{i}+\tfrac{1}{2}%
\epsilon ^{ijk}B_{i}\mathbf{e}_{j}\wedge \mathbf{e}_{k}=\mathbf{i}\frac{%
\mathbf{e}_{0}\wedge \mathbf{x}}{r^{3}}=-\frac{x^{1}}{r^{3}}\mathbf{e}%
_{2}\wedge \mathbf{e}_{3}+\frac{x^{2}}{r^{3}}\mathbf{e}_{1}\wedge \mathbf{e}%
_{3}-\frac{x^{3}}{r^{3}}\mathbf{e}_{1}\wedge \mathbf{e}_{2}  \label{F3}
\end{equation}%
where%
\begin{equation}
r=\left[ \left( x^{1}\right) ^{2}+\left( x^{2}\right) ^{2}+\left(
x^{3}\right) ^{2}\right] ^{1/2}.
\end{equation}%
The field strengths are found to be%
\begin{equation}
\mathbf{E}=0\mbox{\qquad}\mathbf{B}=-\frac{1}{r^{3}}\left(
x^{1},x^{2},x^{3}\right)   \label{F7}
\end{equation}%
as in (\ref{BGLM11}). Since $x\cdot U$ is a scalar in four dimensions, (\ref%
{G2}) provides the extra term in the commutation relation for velocity as%
\begin{equation}
\Delta _{1}=\dfrac{i\hbar }{m}\frac{\left( x\cdot U\right) }{R\left(
x\right) }\mathbf{i}\left( D\wedge x\right) .
\end{equation}%
We may split this expression into rotations and boosts as%
\begin{eqnarray}
\Delta _{1} &=&\dfrac{i\hbar }{m}\frac{x^{0}}{R\left( x\right) }\mathbf{i}%
\left[ \left( D^{0}\mathbf{e}_{0}+\mathbf{D}\right) \wedge \left( x^{0}%
\mathbf{e}_{0}+\mathbf{x}\right) \right]  \\
&=&\dfrac{i\hbar }{m}\frac{x^{0}}{R\left( x\right) }\left[ \left(
D^{0}x^{i}-x^{0}D^{i}\right) \mathbf{i}\left( \mathbf{e}_{0}\wedge \mathbf{e}%
_{i}\right) +\mathbf{i}\left( \mathbf{D}\wedge \mathbf{x}\right) \right] ,
\end{eqnarray}%
and using 
\begin{equation}
\mathbf{i}\left( \mathbf{e}_{\mu }\wedge \mathbf{e}_{\nu }\right) =\epsilon
_{\mu \nu \lambda \rho }g^{\lambda \sigma }g^{\rho \zeta }\mathbf{e}_{\sigma
}\wedge \mathbf{e}_{\zeta },
\end{equation}%
we find%
\begin{equation}
\Delta _{1}=\dfrac{i\hbar }{m}\frac{x^{0}}{R\left( x\right) }\left[ \tfrac{1%
}{2}\epsilon ^{0ijk}\left( D_{0}x_{i}-D_{i}x_{0}\right) \mathbf{e}_{j}\wedge 
\mathbf{e}_{k}+\left( D_{i}x_{j}-D_{j}x_{i}\right) \mathbf{e}_{0}\wedge 
\mathbf{e}_{k}\right] .  \label{exp1}
\end{equation}%
The component commutation relations for velocity with the O(3,1) generators
can be read off as%
\begin{equation}
\left[ \dot{x}^{i},\tilde{M}^{jk}\right] =i\hbar \left( {g}^{ij}{\dot{x}}^{k}%
{-g}^{ik}{\dot{x}}^{j}\right) -\dfrac{i\hbar }{m}\epsilon ^{ijk}\frac{%
x^{0}x_{0}}{R\left( x\right) }\mbox{\qquad}\left[ \dot{x}^{0},\tilde{M}^{jk}%
\right] =0
\end{equation}%
and%
\begin{equation}
\left[ \dot{x}^{i},\tilde{M}^{0k}\right] =i\hbar \left( {g}^{i0}{\dot{x}}^{k}%
{-g}^{ik}{\dot{x}}^{0}\right) -\dfrac{i\hbar }{m}\epsilon ^{ijk}\frac{%
x^{0}x_{j}}{R\left( x\right) }\mbox{\qquad}\left[ \dot{x}^{0},\tilde{M}^{0k}%
\right] =i\hbar {g}^{00}{\dot{x}}^{k},
\end{equation}%
which, under the restriction \mbox{$x\cdot U=x^{0}=0$} of (\ref{4-2}),
become the closed relations. Similarly, applying the restriction (\ref%
{4-1}) to (\ref{finally2}), the commutation relations among the generators
in component form becomes%
\begin{equation}
\left[ {\tilde{M}^{\mu \nu },\,\tilde{M}^{\lambda \rho }}\right] =i\hbar
\left\{ {g^{\mu \lambda }\tilde{M}^{\nu \rho }-g^{\mu \rho }\tilde{M}^{\nu
\lambda }-g^{\nu \lambda }\tilde{M}^{\mu \rho }+g^{\nu \rho }\tilde{M}^{\mu
\lambda }}\right\} +\Delta _{2}^{\mu \nu \lambda \rho }
\end{equation}%
where%
\begin{equation*}
\Delta _{2}^{\mu \nu \lambda \rho }=2i\hbar \frac{x^{2}}{R\left( x\right) }%
\left[ g_{0}^{\mu }\epsilon ^{\nu \delta \lambda \rho }x_{\delta
}-g_{0}^{\nu }\epsilon ^{\mu \delta \lambda \rho }x_{\delta }\right] .
\end{equation*}%
Dividing the generators, we find that the algebra of the boosts is broken%
\begin{eqnarray}
\left[ {\tilde{M}^{0j},\,\tilde{M}^{\lambda \rho }}\right]  &=&i\hbar
\left\{ {g^{0\lambda }\tilde{M}^{j\rho }-g^{0\rho }\tilde{M}^{j\lambda
}-g^{j\lambda }\tilde{M}^{0\rho }+g^{j\rho }\tilde{M}^{0\lambda }}\right\}
+\Delta _{2}^{0j\lambda \rho }  \notag \\
&=&i\hbar \left\{ {g^{0\lambda }\tilde{M}^{j\rho }-g^{0\rho }\tilde{M}%
^{j\lambda }-g^{j\lambda }\tilde{M}^{0\rho }+g^{j\rho }\tilde{M}^{0\lambda }}%
\right\} +\frac{2i\hbar x^{2}}{R\left( x\right) }\epsilon ^{j\delta \lambda
\rho }x_{\delta }
\end{eqnarray}%
while the rotation generators obey closed relations%
\begin{eqnarray}
\left[ {\tilde{M}^{ij},\,\tilde{M}^{\lambda \rho }}\right]  &=&i\hbar
\left\{ {g^{i\lambda }\tilde{M}^{j\rho }-g^{i\rho }\tilde{M}^{j\lambda
}-g^{j\lambda }\tilde{M}^{i\rho }+g^{j\rho }\tilde{M}^{i\lambda }}\right\}
+\Delta _{2}^{ij\lambda \rho }  \notag \\
&=&i\hbar \left\{ {g^{i\lambda }\tilde{M}^{j\rho }-g^{i\rho }\tilde{M}%
^{j\lambda }-g^{j\lambda }\tilde{M}^{i\rho }+g^{j\rho }\tilde{M}^{i\lambda }}%
\right\} 
\end{eqnarray}%
Since only those Lorentz generators that leave the vector $U=\mathbf{e}_{0}$
invariant enjoy the closed commutations relations, this choice of vector $%
U$ restores the closed algebra for the O(3) rotation generators, but not
for the boost generators. Thus, we may understand the nonrelativistic result
as equivalent to the maximal relativistic result for this choice of unit
vector.

We may construct a different kind of solution by choosing $U=\mathbf{e}_{3}$
along the z-axis. Then, from%
\begin{equation}
F\left( x\right) =\frac{x\wedge \mathbf{e}_{3}}{\rho ^{3}}=\frac{x^{0}}{\rho
^{3}}\mathbf{e}_{0}\wedge \mathbf{e}_{3}+\frac{x^{1}}{\rho ^{3}}\mathbf{e}%
_{1}\wedge \mathbf{e}_{3}+\frac{x^{2}}{\rho ^{3}}\mathbf{e}_{2}\wedge 
\mathbf{e}_{3}  \label{F4}
\end{equation}%
we find the field 
\begin{equation}
W\left( x\right) =E^{i}\mathbf{e}_{0}\wedge \mathbf{e}_{i}+\tfrac{1}{2}%
\epsilon ^{ijk}B_{i}\mathbf{e}_{j}\wedge \mathbf{e}_{k}=\mathbf{i}\frac{%
x\wedge \mathbf{e}_{3}}{\rho ^{3}}=\frac{x^{0}}{\rho ^{3}}\mathbf{e}%
_{1}\wedge \mathbf{e}_{2}+\frac{x^{1}}{\rho ^{3}}\mathbf{e}_{0}\wedge 
\mathbf{e}_{2}-\frac{x^{2}}{\rho ^{3}}\mathbf{e}_{0}\wedge \mathbf{e}_{1},
\label{4d}
\end{equation}%
where%
\begin{equation}
\rho =\left[ -\left( x^{0}\right) ^{2}+\left( x^{1}\right) ^{2}+\left(
x^{2}\right) ^{2}\right] ^{1/2}  \label{F5}
\end{equation}%
generalizes the spacial separation in the action-at-a-distance problems of
nonrelativistic mechanics in the subspace%
\begin{equation}
x=\left( x^{0},x^{1},x^{2},0\right) \in x^{\mathbf{e}_{3}}=\left\{
x~|~x\cdot \mathbf{e}_{3}=0\right\}  \label{F8}
\end{equation}%
invariant under the O(2,1) subgroup of the full Lorentz group. The field
strengths are 
\begin{equation}
\mathbf{E}=\frac{1}{\rho ^{3}}\left( -x^{2},x^{1},0\right) \mbox{\qquad}%
\mathbf{B}=\frac{1}{\rho ^{3}}\left( 0,0,x^{0}\right) .  \label{F6}
\end{equation}%
In this case, subject to the restriction (\ref{F8}), the extra terms in the
velocity relations are 
\begin{eqnarray}
\Delta _{1} &=&\dfrac{i\hbar }{m}\frac{x^{3}}{R\left( x\right) }\mathbf{i}%
\left( D\wedge x\right) =0 \\
\Delta _{2}^{\mu \nu \lambda \rho } &=&2i\hbar \frac{x^{2}}{R\left( x\right) 
}\left[ g_{3}^{\mu }\epsilon ^{\nu \delta \lambda \rho }x_{\delta
}-g_{3}^{\nu }\epsilon ^{\mu \delta \lambda \rho }x_{\delta }\right] ,
\end{eqnarray}%
so that closed commutation relations hold for the velocities and among
the O(2,1) generators $\tilde{M}^{01}$, $\tilde{M}^{02}$, and $\tilde{M}%
^{12}=\tilde{L}_{3}$, while the algebra of the generators is broken by field
dependent terms for the boost $\tilde{M}^{03}$ and the rotations $\tilde{M}%
^{31}=\tilde{L}_{2}$ and $\tilde{M}^{23}=\tilde{L}_{1}$.

\subsubsection{Dual field in higher dimensions}

In $N>4$, the multivector $U$ has rank $N-3>1$, and we may gain insight into
this case by partitioning the $N$-dimensional space into the usual
4-dimensional spacetime and $N-3$ `extra dimensions' through%
\begin{equation}
U=n\wedge \mathbf{e}_{4}\wedge \cdots \wedge \mathbf{e}_{N-1}=n\wedge \tilde{%
U}  \label{i1}
\end{equation}%
where 
\begin{equation}
n=n^{\mu }\mathbf{e}_{\mu }\;,\;\mu =0,1,2,3\mbox{\qquad}\tilde{U}=\mathbf{e}%
_{4}\wedge \cdots \wedge \mathbf{e}_{N-1}.
\end{equation}%
Under this partition, (\ref{G3}) becomes%
\begin{equation}
F\left( x\right) =\frac{x\wedge n\wedge \tilde{U}}{R\left( x\right) }
\end{equation}%
subject to the restriction%
\begin{equation}
x\left( \tau \right) \in x^{U}=\left\{ x~|~x\cdot U=x\cdot \left( n\wedge 
\tilde{U}\right) =\left( x\cdot n\right) \tilde{U}-n\wedge \left( x\cdot 
\tilde{U}\right) =0\right\} .  \label{i6}
\end{equation}%
We choose 
\begin{equation}
x\left( \tau \right) =x^{\mu }\mathbf{e}_{\mu }=\left(
x^{0},x^{1},x^{2},x^{3},0,\cdots ,0\right)  \label{i5}
\end{equation}%
so that (\ref{i6}) becomes%
\begin{equation}
x\left( \tau \right) \in x^{U}=\left\{ x~|~\left( x\cdot n\right) =0\right\}
\label{i7}
\end{equation}%
and from (\ref{replacement}) the general solution is%
\begin{equation}
W\left( x\right) =\mathbf{i}\frac{x\wedge n\wedge \tilde{U}}{R\left(
x\right) }.  \label{i8}
\end{equation}%
By similarly partitioning the unit pseudoscalar into the usual 4-dimensional
spacetime and $N-3$ `extra dimensions'%
\begin{equation}
\mathbf{i}=\mathbf{e}_{0}\wedge \cdots \wedge \mathbf{e}_{N-1}=\left( 
\mathbf{e}_{0}\mathbf{e}_{1}\mathbf{e}_{2}\mathbf{e}_{3}\right) \left( 
\mathbf{e}_{4}\cdots \mathbf{e}_{N-1}\right) =\mathbf{i}_{4}~\tilde{U}
\label{i9}
\end{equation}%
the general field becomes%
\begin{equation}
W\left( x\right) =\mathbf{i}_{4}~\tilde{U}\frac{x\wedge n\wedge \tilde{U}}{%
R\left( x\right) }=\mathbf{i}_{4}\frac{x\wedge n}{R\left( x\right) }\tilde{U}%
^{2}=\left[ g_{44}\cdots g_{N-1,N-1}\left( {-1}\right) ^{\frac{\left( {N-4}%
\right) {\left( {N-5}\right) }}{2}}\right] \mathbf{i}_{4}\frac{x\wedge n}{%
R\left( x\right) }  \label{i10}
\end{equation}%
which we recognize as the solution (\ref{4-3}), up to a sign. For this
solution, the extended generators of O($N-1$,1) satisfy the closed
commutation relations, up to extra terms of the form%
\begin{eqnarray}
\Delta _{2} &=&i\hbar \frac{\mathbf{i}x^{2}}{R\left( x\right) }\left\{
\left( D^{\left( 1\right) }\wedge x\right) \wedge \left[ D^{\left( 2\right)
}\cdot \left( n\wedge \tilde{U}\right) \right] -\left( D^{\left( 2\right)
}\wedge x\right) \wedge \left[ D^{\left( 1\right) }\cdot \left( n\wedge 
\tilde{U}\right) \right] \right\}  \notag \\
&=&i\hbar \frac{\mathbf{i}x^{2}}{R\left( x\right) }\left( D^{\left( 1\right)
}\wedge x\right) \wedge \left[ \left( D^{\left( 2\right) }\cdot n\right) 
\tilde{U}-n\wedge \left( D^{\left( 2\right) }\cdot \tilde{U}\right) \right] 
\notag \\
&&-i\hbar \frac{\mathbf{i}x^{2}}{R\left( x\right) }\left( D^{\left( 2\right)
}\wedge x\right) \wedge \left[ \left( D^{\left( 1\right) }\cdot n\right) 
\tilde{U}-n\wedge \left( D^{\left( 1\right) }\cdot \tilde{U}\right) \right] 
\notag \\
&=&i\hbar \frac{\mathbf{i}x^{2}}{R\left( x\right) }\left[ \left( D^{\left(
2\right) }\cdot n\right) \left( D^{\left( 1\right) }\wedge x\right) \wedge 
\tilde{U}-\left( D^{\left( 1\right) }\wedge x\right) \wedge n\wedge \left(
D^{\left( 2\right) }\cdot \tilde{U}\right) \right]  \notag \\
&&-i\hbar \frac{\mathbf{i}x^{2}}{R\left( x\right) }\left[ \left( D^{\left(
1\right) }\cdot n\right) \left( D^{\left( 2\right) }\wedge x\right) \wedge 
\tilde{U}-\left( D^{\left( 2\right) }\wedge x\right) \wedge n\wedge \left(
D^{\left( 1\right) }\cdot \tilde{U}\right) \right] .
\end{eqnarray}%
As in $N=4$, the extra terms vanish for three of the O(3) or O(2,1)
generators --- the three are determined by the conditions%
\begin{equation}
D^{\left( 1\right) }\cdot \tilde{U}=D^{\left( 2\right) }\cdot \tilde{U}=0%
\mbox{\qquad}\Rightarrow \mbox{\qquad}D^{\left( 1,2\right) }=D^{\left(
1,2\right) \mu }\mathbf{e}_{\mu }\;~\;,\;\;\;\mu =0,1,2,3
\end{equation}%
and%
\begin{equation}
D^{\left( 1\right) }\cdot n=D^{\left( 2\right) }\cdot n=0.
\end{equation}%
The examples in the previous section are recovered by the choices $n=\mathbf{%
e}_{0}$ and $n=\mathbf{e}_{3}$.

\section{Conclusion}

We have seen that in the presence of the field 
\begin{equation}
W^{\mu \nu }\left( x\right) =\frac{1}{{\left( {N-2}\right) !}}{\epsilon
^{\mu \nu \lambda _{0}\lambda _{1}\cdots \lambda _{N-3}}}F_{\lambda
_{0}\lambda _{1}\cdots \lambda _{N-3}}=\frac{1}{{\left( {N-3}\right) !}}%
\frac{{\epsilon ^{\mu \nu \lambda _{0}\lambda _{1}\cdots \lambda
_{N-3}}x_{\lambda _{0}}U_{\lambda _{1}\cdots \lambda _{N-3}}}}{{R\left(
x\right) }}
\end{equation}%
the extended O($N-1,1$) generators%
\begin{equation}
{\tilde{M}^{\mu \nu }=M^{\mu \nu }+Q^{\mu \nu }=m}\left( x^{\mu }\dot{x}%
^{\nu }-x^{\nu }\dot{x}^{\mu }\right) +x^{\mu }x_{\sigma }W^{\sigma \nu
}-x^{\nu }x_{\sigma }W^{\sigma \mu }-x^{\sigma }x_{\sigma }W{^{\mu \nu }}
\end{equation}%
are constructed in such a way that the commutation relations of the
generators with position ${x^{\mu }}$ and velocity ${\dot{x}^{\mu }}$
satisfy the closed Lie algebra 
\begin{equation}
\left[ {x^{\mu },\,\tilde{M}^{\rho \lambda }}\right] =i\hbar \left( {%
x^{\lambda }g^{\mu \rho }-x^{\rho }g^{\mu \lambda }}\right) \mbox{\qquad}%
\left[ {\dot{x}^{\mu },\,\tilde{M}^{\rho \lambda }}\right] =i\hbar \left( {%
g^{\mu \rho }\dot{x}^{\lambda }-g^{\mu \lambda }\dot{x}^{\rho }}\right) 
\end{equation}%
when the dynamical evolution is restricted to%
\begin{equation}
x\left( \tau \right) \in x^{U}=\left\{ x~|~x^{\lambda _{1}}U_{\lambda
_{1}\lambda _{2}\cdots \lambda _{N-3}}=0\right\} 
\end{equation}%
and%
\begin{equation}
{R\left( x\right) =}\left( x^{2}\right) ^{3/2}.
\end{equation}%
Similarly, and the commutation relations among the generators%
\begin{equation}
\left[ {\tilde{M}^{\mu \nu },\,\tilde{M}^{\lambda \rho }}\right] =i\hbar
\left\{ {g^{\mu \lambda }\tilde{M}^{\nu \rho }-g^{\mu \rho }\tilde{M}^{\nu
\lambda }-g^{\nu \lambda }\tilde{M}^{\mu \rho }+g^{\nu \rho }\tilde{M}^{\mu
\lambda }}\right\} +\Delta _{2}^{\mu \nu \lambda \rho }
\end{equation}%
with%
\begin{equation}
\Delta _{2}^{\mu \nu \sigma \rho }=i\hbar \frac{x^{2}}{R\left( x\right) }%
\frac{1}{\left( N-3\right) !}\epsilon ^{\mu \sigma \rho \zeta \lambda
_{2}\cdots \lambda _{N-3}}x_{\zeta }g^{\nu \lambda _{1}}U_{\lambda
_{1}\lambda _{2}\cdots \lambda _{N-3}}
\end{equation}%
satisfy the closed Lie algebra for the three generators of O(3) or O(2,1)
that leave the subspace $x^{U}$ invariant. We also showed that by
appropriate choice of $U$ it is possible to recover a given four-dimensional
solution in any number of dimensions. In particular, the O(3)-invariant
solution recovers the nonrelativistic case for any $N$. Thus, the solution
can be interpreted as a generalization of the Dirac monopole to $N>4$. The
field strength in the O(2,1)-invariant solution is associated with a
potential of the type 
\begin{equation}
V\left( x\right) \sim \left( -t^{2}+\mathbf{x}^{2}\right) ^{-1/2}
\end{equation}%
which may be seen as a relativistic generalization of the nonrelativistic
Coulomb potential. A solution to the relativistic bound state problem for
the scalar hydrogen atom was found \cite{bound} in the context of the
Horwitz-Piron \cite{H-P} formalism, using a potential of this form. It is
noteworthy, that this form of potential does not generally follow from the
wave equation in $N$-dimensions. The interpretation of the generalized Dirac
monopole and its relationship to the relativistic generalization of the
bound state problem is discussed in a second paper.

\end{document}